%% file: ss.tex
\documentclass{article}
\usepackage{epsfig}

\begin{document}

\title{On free evolution of self gravitating, spherically symmetric waves.~\footnote{Work supported
             by Antorchas, CONICOR, CONICET and Se.CyT, UNC}}

\author{{\sc Mirta S. Iriondo}\thanks{Partially supported by AIT, C\'ordoba,
    Argentina. },  \\
   {\sc Oscar A. Reula}
  \thanks{Member of CONICET.}\\
  {\small FaMAF, Medina Allende y Haya de la Torre,}\\
  {\small Ciudad Universitaria, 5000 C\'ordoba, Argentina}\\
}

\maketitle 

\vspace{-.4in}

\begin{abstract}
  We perform a numerical free evolution of a selfgravitating, spherically 
symmetric scalar field satisfying the wave equation. 
The evolution equations can be written in a very simple form and are 
symmetric hyperbolic in Eddington-Finkelstein coordinates.
The simplicity of the system allow to display and deal with the typical
gauge instability present in these coordinates. 
The numerical evolution is performed with a standard method of lines
fourth order in space and time. The time algorithm is Runge-Kutta while the
space discrete derivative is symmetric (non-dissipative). 
The constraints are preserved under evolution (within numerical errors) 
and we are able to reproduce several known results.

\end{abstract}


\section{Introduction}
\label{sec:int}

In the past decade, due to the results of several numerical experiments, 
there has been a considerable understanding of the 
phenomena of the collapse of a spherically symmetric, selfgravitating scalar 
field. 
In particular universal
scaling properties of the final mass of the black hole were discovered and
then explained. For a review and references see \cite{Gundlach}
The present work is not aimed to continue that line, but rather to study 
instability
problems which are recurrent in many numerical evolutions near black holes
when those evolutions are completely free, that is when the constraints are
not solved for, but remain being satisfied just because the evolution 
equations are solved for. 
There are few cases in which free evolution has
been successful, some of them, 
\cite{Gomez,Garfinkle,Burko_1} use light cone coordinates and in fact some
of the equations being solved could be considered as constraints. 
Others, 
\cite{Sch-Baum_1,Sch-Baum_2}, uses a symmetric hyperbolic formulation and use the freedom 
still available in that setting to suppress the main instability found there 
and so reach a stable propagation, alas first order numerical dissipation is 
used in that scheme. 
Finally there are conformal space-time methods
\cite{Huebner,Frauendiener} 
where no instability seems to
be present, nor should be very relevant, for only short time evolutions suffices
to cover the whole future domain of dependence.

In this work we use Eddington-Finkelstein coordinates and are able to choose 
equations and variables so that the final system of equations is remarkably 
simple and manifestly symmetric hyperbolic. Such simplicity allows as to
study the nature of the instability and proceed to avoid it.
  
In the next section we introduce the equations and discuss which evolution 
system, among all possible equivalent ones, is suitable for free evolution. 
It turns out that two out of the five characteristics of the system
can be specified at will by modifying the evolution equations. This is done
by adding terms proportional to the constraints. 
These characteristics can even
become complex, signaling the failure of the evolution equations to
be hyperbolic, and so to have a well posed initial value formulation.
The change on the characteristics of the evolution system reflects on
an identical change on the characteristics of the evolution equations
satisfied by the constraint equations. Thus it is possible to understand
that the change in boundary conditions induced by the change on the direction
of the characteristics is just the one needed to ensure the correct 
preservation of the constraint equations under evolution.
Among all possible evolution system we take a specific one which is
both the simplest in terms of computations needed and also on dealing
with constraint preservation.
We then look at
the residual gauge freedom of the Eddington-Finkelstein coordinates, that 
freedom is the only \textsl{dynamics} which is present in the vacuum case, and
is the mode that can generate instabilities. 
After that we discuss the initial-boundary value problem for the specific
evolution system chosen and determine which fields must be given 
in order to have a unique evolution, in particular we discuss a case in
which the characteristics change at the inner boundary according to whether
that boundary is inside or outside the horizon. 
We then discuss the instabilities of the system, due to the simplicity of
the evolution system chosen this task is trivial and we can display explicitly
the instability. We can decide in which cases there will be
problems and the type of them. In particular near the stationary regime, that
is when most of the scalar field has dissipated away or fallen into the 
black hole the system can be made stable using simple initial-boundary values 
for the gauge fields. 

Having completed the analytic study of the equations in 
 section \ref{sec:NM} we discuss the numerical methods employed, look at
the convergence tests and constraint preservation under evolution for
several types of initial-boundary data sets.

In the next section we comment in several results obtained by the simulations,
they do not pretend to be novel or interesting, but are done just to 
confirm the good properties of the evolution system chosen.
They include ringing and tail studies, the relation mass-flux for boundary
data and the relation total mass black hole mass for different types of
collapse.



\section{The equations}
\label{sec:eqns}


We consider the Eddington-Finkelstein coordinate system (\cite{Marsa_Choptiuk}), 
the time dependent spherically symmetric metric is
$$
ds^2=(-\alpha^2+a^2\,\beta^2)\, dt^2+2a^2\beta\, dt\, dr+a^2\,dr^2+r^2b^2\,d\Omega^2 
$$
with $a, b, \alpha$ and $\beta$ being functions of $r$ and $t$, and $d\Omega^2$ 
the metric of the unit-sphere. 
The areal locking condition ($\dot{b}=0$) and 
the condition that the vector $\partial_t-\partial_r$ be null is achieved 
choosing the lapse and the shift as
$$
\beta=\frac{ra\, K^\theta{}_\theta}{1+ra\, K^\theta{}_\theta}\qquad \alpha=\frac{a}{1+ra\,K^\theta{}_\theta}.
$$
where $K^\theta{}_\theta$ is the corresponding coordinate component of the extrinsic curvature
of the constant time hypersurfaces.
Moreover this choice corresponds to the choice  $b=1$, and with it the metric becomes

$$
ds^2=a^2((2\beta-1)\,dt+dr)\,(dt+dr)+r^2\,d\Omega^2. 
$$ 
The full set of evolutions  and constraint equations for the geometric variables 
$(a,K^\theta{}_\theta,K^r{}_r)$ are~\footnote{
Notice some corrections with respect to the equations in \cite{Marsa_Choptiuk}, 
also notice the difference on the choice of gravitational constant.}

\begin{eqnarray}
\partial_t a&=&\partial_r(a\beta)-a^2(1-\beta)\,K^r{}_r\nonumber\\
\partial_t K^\theta{}_\theta&=&\beta\,\partial_rK^\theta{}_\theta+\frac{\partial_r\beta}{ar}+a(1-\beta)\,K^\theta{}_\theta\,(K^r{}_r+2 K^\theta{}_\theta )+\frac{1-\beta}{r^2}\left(a-\frac{1}{a}\right )\nonumber\\
\partial_t K^r{}_r&=&\beta \partial_r K^r{}_r+\frac{\partial^2_r\beta}{a}+\frac{\beta-1}{a}\left[\frac{\partial^2_r a}{a}-\left(\frac{\partial_r a}{a}\right)^2-\frac{2\partial_r a}{ra}\right]+\frac{\partial_r\beta\partial_r a}{a^2}\label{evolI}\\
&&+a(1-\beta)\,K^r{}_r\,(K^r{}_r+2 K^\theta{}_\theta )+\frac{8}{a}\Phi^2 (\beta-1).\nonumber
\end{eqnarray}
and
\begin{eqnarray}
C_a&=&\partial_ra+\frac{1}{2r}(a^3-a)+\frac{a^3 r}{2}\,K^\theta{}_\theta\,(2K^r{}_r+ K^\theta{}_\theta )-2 r a\,(\Phi^2+\Pi^2)=0\nonumber\\
C_{K^\theta{}_\theta}&=&\partial_r K^\theta{}_\theta+\frac{ K^\theta{}_\theta-K^r{}_r}{r}-\frac{4}{a}\Phi\Pi=0\label{constraintI}
\end{eqnarray}
respectively.
While the wave equation for the massless scalar field becomes the first-order (in time) 
system
\begin{eqnarray}
\partial_t \Phi&=&\partial_r \bigg(\beta\Phi+(1-\beta)\Pi\bigg)\nonumber\\
\partial_t \Pi&=&\frac{1}{r^2}\partial_r\left(r^2(\beta\Pi+(1-\beta)\Phi\right)\label{fieldI}
\end{eqnarray}
where $\Pi = \partial_t \phi$, and $\Phi = \partial_r \phi$.

In order to obtain  first order evolution equations, we perform the  change of variables 
$(f, g, h, \chi_+, \chi_-)$
$$
f=K^\theta{}_\theta\, a\, r +1\qquad 
g=\frac{a^2}{f}\qquad 
h=\frac{1}{af}\left(\frac{a}{2}\partial_r f+\partial_r a- a^2 K^r{}_r\right).
$$

$$
\chi_+=\frac{(\Pi+\Phi)r}{2}\qquad \chi_-=\frac{(\Pi-\Phi)r}{2}
$$
with this change the shift becomes $\displaystyle{\beta=\frac{f-1}{f}}$ and  
$$
ds^2=g\left((f-2)\, dt +f\, dr\right )( dr +dt)+ r^2\, d\Omega^2.
$$

Then the full set of evolution and constraint equations  for the metric variables $(f,g,h)$ are 
\begin{eqnarray}
\partial_t f&=&\frac{\partial f}{\partial r}+\frac{g-2+f}{r}\nonumber\\
\partial_t g&=&-\frac{g}{f}\,\frac{\partial f}{\partial r}+\frac{f-2}{f}\,\frac{\partial g}{\partial r}+2 g h-\frac{g(g-2+f)}{r f}\label{evolII}\\
\partial_t h&=&\frac{\partial h}{\partial r}- \frac{1}{2f r}\frac{\partial f}{\partial r}+\frac{g-2}{2\, f h r}\,\frac{\partial g}{\partial r}-\frac{(g-f)h}{fr}-\frac{(g-2+f)}{2\,f r^2}+ 2 (\frac{\chi_++\chi_-}{f r})^2 \nonumber
\end{eqnarray}
and
\begin{eqnarray}
C_f&=&\frac{\partial f}{\partial r}-2(f-2)h+\frac{f+g-2}{r}-4\frac{\chi_+^2}{f r}=0\nonumber\\
C_g&=&\frac{\partial g}{\partial r}-\frac{2g(hf r-\chi_+^2+\chi_-^2)}{f r}=0\label{constraintII}
\end{eqnarray}
while the wave equation for the massless scalar field becomes 
\begin{eqnarray}
\partial_t \chi_+&=&\frac{\partial \chi_+}{\partial r}-\frac{(f-2)\chi_-}{f r}\nonumber\\
\partial_t \chi_-&=&\frac{f-2}{f}\,\frac{\partial \chi_-}{\partial r}+\frac{2\chi_-}{f^2 r} (2r(f-2)h+2-g-f)+\frac{8(\chi_-)\chi_+^2}{f^3 r}-\frac{\chi_+}{r}\label{fieldII}
\end{eqnarray}
where we have already used one of the constraint equations.

Clearly the evolution equations for the scalar field are symmetric hyperbolic
as a subsystem and their characteristics are fixed, they correspond to the
two null directions. 
On the other hand the set of evolution equations for the metric coefficients 
is not unique, for we can add to them terms proportional to the constraint
equations and change some of their characteristics at will.
To study this possibilities we add to the   evolution equations for $f$, $g$, 
and $h$ respectively the following terms 
$(K_{ff}-1) C_f + K_{fg} C_g$, 
$(K_{gf}+\frac{g}{f}) C_f + (K_{gg}-\frac{f-2}{f}) C_g$, 
$(K_{hf}-\frac{g-2}{2frh}) C_f + (K_{hg}+\frac{1}{2fr}) C_g$. 
We get a system with the following principal part matrix,~\footnote{Recall that the principal part matrix is the one form by the coefficients of the derivative terms appearing on the equations.}

\[
 \left( 
    \begin{array}{ccc}
         K_{ff} & K_{fg} &  0 \\
         K_{gf} & K_{gg} &  0 \\
         K_{hf} & K_{hg} &  1
    \end{array}
    \right)
\]

The characteristics of this system are given by the one forms 
$(1,1)$, $(1,\lambda_+)$, and $(1,\lambda_-)$, where 
$$
\lambda_{\pm} = (K_{ff} + K_{gg} \pm \sqrt{(K_{ff}-K_{gg})^2 + 4 K_{fg} K_{gf}})/2.
$$
The first one corresponds to the propagation of $h$ and it does so along 
the null incoming direction. It is the only fix characteristic. 
The other
two can take any values, for instance if we take $K_{fg}=K_{gf}=0$ then the
characteristics are $(1,K_{ff})$, and $(1,K_{gg})$, thus choosing these two 
values we can prescribe any propagation direction we please, they 
can be incoming, or outgoing time-like, null or even space-like. 
Even more, choosing 
$K_{ff}=K_{gg}=0$, $K_{fg}=-K_{gf}\neq 0$ the characteristics become imaginary and
so the system is not even hyperbolic! So we see that the different ways of writing Einstein's equations are only 
equivalent in the sense that solutions of one are solutions of the others, 
but some of their evolution equations do not even have a well posed initial 
value problem. 
We have tried to evolve numerically a system as the last one 
(with $K_{fg}=-K_{gf}= 1$) and as expected it is completely unstable, 
giving an explosive blow up of higher wave number modes. 
To avoid the fact that no
good boundary conditions are known for non-symmetric hyperbolic systems
these equations were tested using periodic boundary conditions and the initial 
data did not satisfy the constraint equations.
But the instability does not seems to be sensitive to the failure of the
constraint equations to be fulfilled nor to the particular boundary values chosen for
those instabilities appear locally at every point of the evolution hypersurface.

Thus, to start with, we must choose coefficients
such that the system is hyperbolic. 
Since the coefficients $K_{hf}$,  $K_{hg}$
do not alter the characteristics we shall take them to vanish, thus 
simplifying a bit the numerical computations.
The sign of the characteristics at the boundary points determine whether or
not values for the different fields must be given or not as boundary 
conditions. 
But not all these values are free, for they must be consistent with 
constraint propagation. 
In order for the constraints to remain enforced along evolution we must make 
sure that the resulting boundary conditions for the evolution equations 
obeyed 
for the constraints ensure that they have a unique solution, and it is the 
trivial one. 
It follows trivially from the form of the constraints that the characteristic
matrix for the constraints evolution equations is:

 \[
 \left( 
    \begin{array}{ccc}
         K_{ff} & K_{fg}  \\
         K_{gf} & K_{gg} 
    \end{array}
    \right)
\]
So that the propagation directions for them coincide with the propagation
directions of the pair $(f,g)$. This fact seems to imply that the freedom
in choosing part of the pair $(f,g)$ at the boundaries is just the one 
needed for getting maximally dissipative boundary conditions for the 
constraint system, thus making sure they propagate correctly. 
Unfortunately even for this simple case a rigorous proof of this has proven 
be difficult~\footnote{The main difficulty arises when the $K$ coefficients 
depend on the metric components}, 
and the resulting conditions are complicated to implement numerically. 
Thus it seems that the best one can do retaining simplicity is
to choose coefficients so that the characteristics for the pair $(f,g)$ are
incoming or tangent in both boundaries. 
This way no boundary condition is needed for the pair $(f,g)$ and therefore no
boundary condition must be satisfied by the constraints, thus, there 
is no danger that the constraint would cease to be satisfied during evolution.
The simplest case of all is when 
both are tangent, and this is the case for instance if we take all
coefficients to vanish. With this choice the evolution system reduces to,

\begin{eqnarray}
\partial_t f&=&2(f-2)h+\frac{4\chi_+^2}{rf}\nonumber\\
\partial_t g&=&2gh-\frac{2g\chi_+^2}{f r} + \frac{2g(f-2)\chi_-^2}{r f^2}\label{evolIII}\\
\partial_t h&=&\frac{\partial h}{\partial r}-\frac{(g-2)\chi_+^2}{f^2r^2}+\frac{4\chi_+\chi_-}{f^2 r^2}+\frac{g\chi_-^2}{f^2 r^2}\nonumber
\end{eqnarray}

And similarly the evolution equations for the constraints are just ordinary
differential equations, 


\begin{eqnarray}
\partial_t C_f&=&\frac{2}{r\,f^2}(h\,f^2r-2\chi_+^2)\,C_f\label{evolconstI}\\
\partial_t C_g&=&\frac{2 g}{r\, f^3}(f \chi_+^2-f \chi_-^2+\chi_-^2) \,C_g+\frac{2 }{r\, f^2}(h r f^2-  f\chi_+^2-2 \chi_-^2+f\chi_-^2)\,C_g \label{evolconstII}
\end{eqnarray}

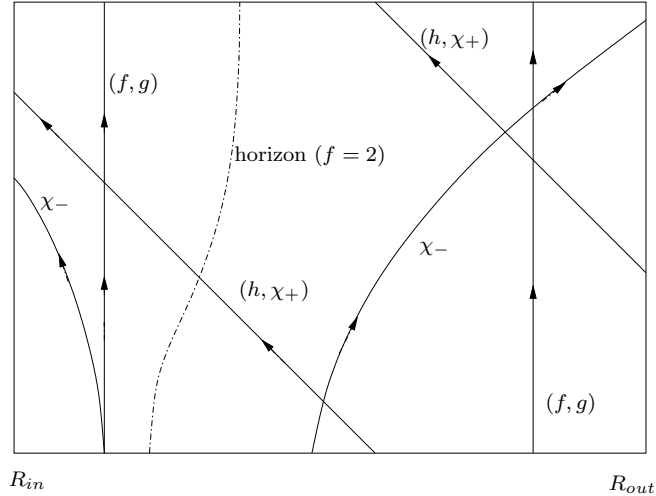
\begin{figure}[h]
  \begin{center}
    \input{char.pstex_t}
    \caption{Integration region with characteristics lines}
    \label{fig:char}
  \end{center}
\end{figure}

Since they are homogeneous ordinary differential equations the constraints 
will remain to be satisfied along evolution if they are initially satisfied.
No extra condition is need at (nor can be imposed on) the boundaries.
This is the system we shall study in detail from now on. Note that this system
not only is symmetric hyperbolic, but it is also diagonal. 
In fact we reach such a simple system when looking for variables which would
give diagonal systems.
In figure \ref{fig:char} we show all characteristics of the chosen system.

\vspace{2mm}



\subsection{The Gauge Freedom}
\label{subsec:GF}

The chosen lapse and shift functions depend on the dynamical variables and 
therefore do not fix completely the gauge.
To display this freedom we introduce  the null coordinate $u=(t+r)$~\footnote{We thanks M. Tiglio for pointing to us this way to proceed.}.
In the $(u,r)$ coordinates the metric becomes
$$
ds^2 = g((f-2)du + 2dr)du +  r^2\, d\Omega^2, \label{met}
$$
For Schwarzschild the choice $g=1$, $f-2 = -(1-2M/r)$ gives the standard
form of the metric.
If we choose some other coordinate $\tilde{u} = \tilde{u}(u)$ the functional form remains 
the same if 
$\tilde{g}= g/\frac{d\tilde{u}}{du}$ 
and 
$\frac{\tilde{f}-2}{\tilde{}g} = \frac{f-2}{g}$.

The invariant ratio $\frac{f-2}{g}$ is related to the mass function, indeed,

$$
m = \frac{r}{2}(1 + \frac{f-2}{g}) 
$$
is the usual form of the local mass for spherical symmetric space-times.

Another invariant is the surface where $f$ takes the value $2$, this can 
be seen to be an apparent horizon.

From the first constraint equation, (\ref{constraintII}), it follows that
under this gauge transformation, in the vacuum case,

$$
\tilde{h} = h + \frac{1}{2}\frac{\frac{d\tilde{u}}{du}}{\frac{d^2\tilde{u}}{du^2}}
$$

To fix completely the gauge we must uniquely define the coordinate $u$, 
this can be done by choosing their values at the initial time slice and at 
the outer boundary of the integration region.
We shall do this by choosing the values for $h$ at those points.


\subsection{Initial Boundary Value Problem}
\label{subsec:IBV}


We are interested in performing a time evolution between two fixed 
(areal) radiae, $R_{in}$, $R_{out}$, starting at a given initial surface, $t=t_0$.
Thus we have to see what initial-boundary values can be prescribed.

With the standard choice of equations (\ref{fieldII},\ref{evolIII}) it is 
clear that we need to give initial values for the fields $(f,g,h,\chi_+,\chi_-)$.
Since they must satisfy the constraint not all initial data can be freely
specified. 
We have taken a set of initial data sets for which the constraints
are easily solved. We give an arbitrary initial value for $\chi_+$ and
solve
$$
\frac{\partial m}{\partial r} = {\chi_+}^2,\qquad m|_{r=r_{in}}=m_0,
$$
Then the following choice is a solution of both constraint equations,

\begin{eqnarray}
    \chi_- &:=& 0, \nonumber\\
  g &:=& 1, \label{eq:initial_data} \\
  f &:=& 1 + 2m/r \nonumber\\
  h &:=& \frac{{\chi_+}^2}{rf} \nonumber\\
\end{eqnarray}

In our numerical simulations we have taken $\chi_+$ to be a Gaussian or a 
power of r times an exponential decay, so that the mass integration could 
be done exactly. The integration constant for $m$, the mass at $R_{in}$, can
take any value, in general we have taken either~\footnote{
Due to the scaling properties of the system there is no loss of generality 
in choosing only just these two values.} 
$m|_{R_{in}} = 1$ or $m|_{R_{in}} = 0$

We now discuss the boundary conditions (see figure (\ref{fig:char}).
If the inner boundary is inside the apparent horizon (which is located at
$f=2$) there is no incoming 
characteristic at it and so no boundary condition is needed (nor can be prescribed) 
there. 
In fact in this case this boundary is actually space-like.
If the inner boundary is outside the horizon there is one characteristic incoming into 
the integration region, namely the one for $\chi_-$, so
we have to prescribe some value for it. To allow for an apparent horizon 
marching across the boundary the numerical code checks for the value of 
$f$ at the boundary and, depending whether its value is smaller or bigger 
than 2, imposes a boundary condition for $\chi_-$. 
When needed we have normally used a null incoming radiation condition, 
($\chi_-|_{R_{in}}=0$) but it is clear
than in that case there is no natural boundary condition which can mimic
the physical collapse one is trying to simulate.

At the outer boundary, and provided the horizon is towards the inside of
that boundary ($f|_{R_{outer}} < 2$), there are two incoming characteristics,
one corresponding to $\chi_+$ the other to $h$. We can prescribe arbitrary
values for them there, and we have performed several runs prescribing 
either incoming scalar field radiation or pure gauge modes.  

With these initial-boundary value conditions the problem is well posed and
so, given smooth data we obtain a smooth solution valid for a finite time 
interval.


\subsection{Analytic Instabilities}
\label{sec:AI}


The evolution system is well posed, so there are no instabilities of the
explosive type, that is those whose rate grows unbounded with the frequency, 
rather the expected instabilities have an exponent with bounded positive 
real part. 
These instabilities essentially affect the pair $(f,g)$ which controls
the characteristics of the system, if they are present, and act for a
long enough time they can ruin the hyperbolicity of the system --by making
some of the propagation speeds to diverge--
and so its well posedness.
But much before that happens one of the characteristic speeds
grows so much that the system becomes numerically unstable due to violations 
of the Courant-Friedrich stability condition.

For simplicity we analyze the instability for the vacuum case.
In that case the equations are:

\begin{eqnarray}
\partial_t f&=&2(f-2)h \nonumber\\
\partial_t g&=&2gh \label{evol_vacuum}\\
\partial_t h&=&\frac{\partial h}{\partial r} \nonumber
\end{eqnarray}

Given initial data for $(f,g,h)$ and boundary data at the outer boundary for $h$, we 
can first solve for $h$,  $h(t,r)=h_b(t+r-R_{out})$ and then for $(f,g)$. We have,

\begin{eqnarray}
  f(t,r)-2 &=& (f_0(r)-2)e^{2\int_0^t h_b(\tilde{t}+r-R_{out})d\tilde{t}} \nonumber\\
  g(t,r) &=& g_0(r)e^{2\int_0^t h_b(\tilde{t}+r-R_{out})d\tilde{t}} \nonumber
\end{eqnarray}

Thus we see that $g$ remains always positive, and the horizon --the radius 
at which $f(t,r)=2$ remains fixed. But the equal time surfaces tilt, and so
if $h$ takes big positive values, $f$, and $g$ grow exponentially and we
have an analytical instability, which in numerical simulations would result in 
a numerical one if no appropriate methods are used. 
If a scalar field is present, no matter how small, the tilt of the equal time
surfaces would increase the propagation speed of the outgoing modes, 
$v = \frac{f-2}{f}$. 
At points inside the horizon ($f_0-2 > 0$) and so there $f-2$ grows exponentially.
The propagation speed $v$ goes to $1$. At points outside the horizon ($f_0-2 < 0$)
and so there $f$ eventually becomes negative. 
Notice that when $f$ vanish the equal time hypersurface becomes null and 
so the time evolution of all outgoing modes (in this case only the one
from the scalar field) is ill posed in this gauge.

On the other hand if $h$ becomes large but negative, $f \to 2$ and now
the hypersurface of constant $r$ approach null surfaces. 
In that case the evolution still can be done, but the loss of
accuracy is very important, even for moderate values of $h$.
 
We need therefore to keep $h$ moderately small during evolution and 
propagating inwards without any important {\sl residual part} due to
the numerical method. This can be achieved in many cases by just 
prescribing the initial-boundary conditions for $h$ so that it starts 
small. 
There are nevertheless situations for which $h$ becomes big and
the simulation runs into the instabilities, this is so, for instance, if
one takes boundary data for the incoming component of the scalar field
with a very big amplitude (final masses of order $10$). 
In that case the $h$ generated along the evolution of the scalar field
is important and the systems becomes unstable. 
On the other hand, for situations describing the final
stages of collapse, namely after most of the scalar field has fallen into
the black hole and the geometry has settled in a near stationary regime the
instabilities do not show up. We have tested that with evolutions lasting
for several thousand masses.


\section{Numerical Methods}
\label{sec:NM}


To perform the simulations we use a standard fourth order method of lines. 
The space derivatives discretization is done with standard fourth order
(we have also used second order discretizations with similar results) discretized 
derivatives which are one sided at the boundaries. 
We have used symmetric and non-symmetric, but full fourth order, ones. 
The symmetric ones, \cite{Kreiss, Strand}, are guaranteed to be stable, 
but are only second or third order accurate at the last points of the 
grid near the boundary. 
This accounts, for instance, for spikes at the error on the constraint 
propagation check, some of these errors sometimes propagate inwards, but 
stay bounded and diminish as the grid spacing is decreased.
The non-symmetric discretizations used were full fourth order 
(but with bigger errors near the boundary) and they give
smaller errors on constraints evolution near the boundaries.
But there is no guaranty that the scheme would be stable. 
Most of the calculations, including all the results and test given in this
paper, were done with the symmetric ones. Their symmetry implies they are norm
preserving and so they do not have any dissipation.
Fourth order centered differences approximations introduce hight frequency modes 
with backwards propagations speeds (group velocities) which are 5/3 bigger 
than the exact speeds. 
These modes become important when analyzing the tail behavior of 
the scalar field, for there the values of the scalar field become so small
that the contribution from them becomes important. 
The higher than light propagation of them implies that boundary 
effects take place before they should. These high frequency modes become smaller 
as the grid size is decreased, thus tail decay studies are possible if 
enough grid points are
included, but this should be very costly in higher dimensions.
For these studies it seems better to use centered difference schemes of 
second order, for they have a maximum group velocity of one.

Time discretization is done using the standard fourth order Runge-Kutta 
method.

We have done three types of simulations: The first type of simulation has
no-incoming boundary conditions ($\chi_+=h=0$ at the outer boundary) 
and an initial data satisfying (\ref{eq:initial_data}).
One starts with a solution representing an inner black hole (usually of 
unit mass) and certain amount of infalling scalar field. The evolution
should result in a bigger black hole and some amount of scalar
field going outwards. Thus a ringing and latter a power law tail should
be observed in the outgoing component of the scalar field, $\chi_-$. 

The second type are vacuum solutions representing pure gauge dynamics.
They have Schwarz\-schild or Minkowski initial data and the dynamics is
created by allowing incoming gauge ($h|_{R_{out}} = h^b(t)$)
to enter the outer boundary.
The evolution should have some dynamics and latter on, if the incoming 
gauge has a finite duration, should relax into a static black hole in a
different gauge, through the simulation the mass should stay constant,
as should the horizon position. 

The third type of data starts with initial data corresponding to a 
Schwarzschild black hole of unit mass and latter on the evolution an
incoming scalar field mode is injected into the system from the outer
boundary during a finite amount of time ($\chi_+|_{R_{out}} = \chi^b_+(t)$. 
The evolution should create 
a bigger black hole and some outgoing radiation should be seen.

We shall discuss the results for these three type of simulations in the
next section.
 

\subsection{Convergence Test}
\label{subsec:CT}

\begin{figure}[ht]
\noindent
\begin{minipage}[t]{.46\linewidth}
\centering \epsfig{angle=-90,width=6.5cm, figure=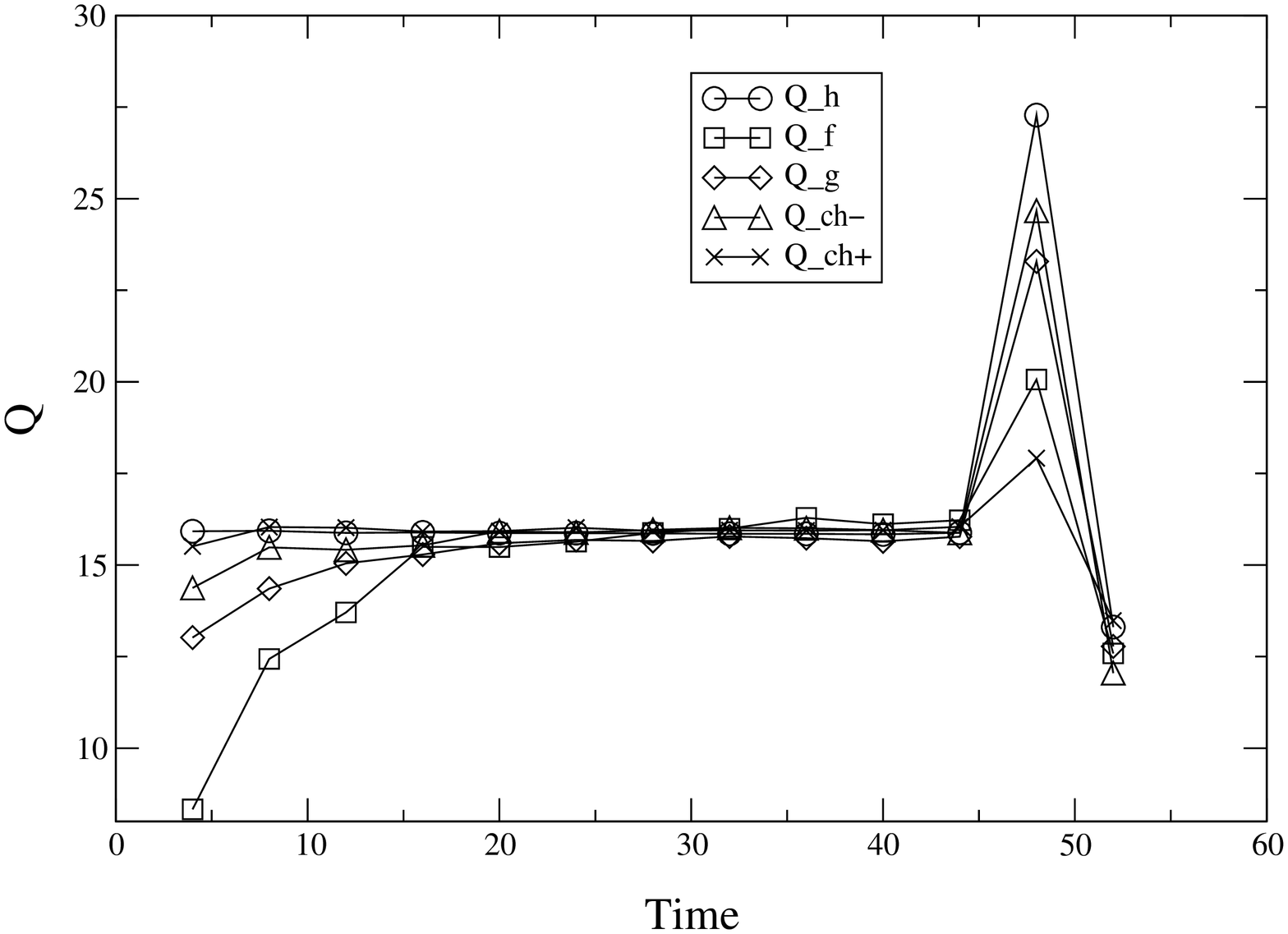}
\caption{$Q$ factor for $L^2$ norms of type I data}
\label{fig:48L2}\hspace{2mm}
\end{minipage}\hfill
\begin{minipage}[t]{.46\linewidth}
\centering \epsfig{angle=-90,width=6.5cm, figure=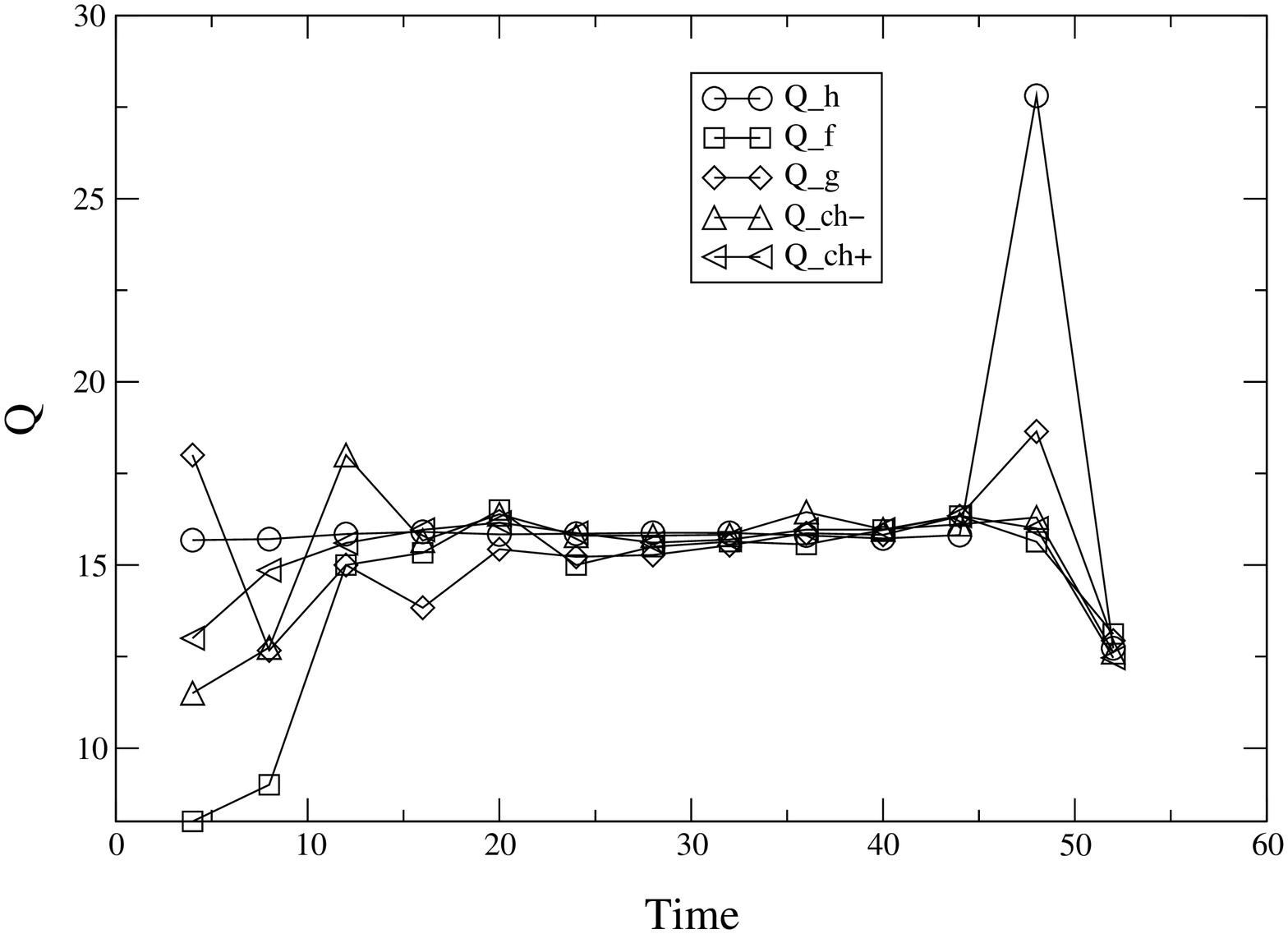}
\caption{$Q$ factor for $L^{\infty}$ norms of type I data}
\label{fig:48Linf}
\end{minipage}\hfill
\end{figure}

We have perform convergence tests for the three types of simulations described above. 
In general it is difficult to analyst the convergence order in detail
due to the fact that the space discrete derivative we use is not of the
same order near the boundary (third or second order) rather than fourth.
This gives rise to Q values different than the corresponding to fourth order
(Q=16). 
Typically after boundary effects become important 
(that is when the fields reach the boundary) we get values over 12, but some 
times they drop to about 4 when the bulk of the solution is near the boundary.
In figure: \ref{fig:48L2}, and \ref{fig:48Linf} we give values for the 
Q factors obtained using the $L^2$ and $L^{\infty}$ norms respectively,
for a run with homogeneous boundary conditions and a Gaussian peak of the 
ingoing component of the scalar field midway in the integration region.
This run has an initial black hole mass $M_{Initial}=1$,
a total mass, $M_{Total}=8.90$ and a final black hole mass of $M_{Final} = 8.67$
so it is well into the nonlinear regime.
The peak observed at the end of the integration period is due to boundary
effects, after that the values of the different Q's go to values above 12,
but never return to 16. 

\vspace{5mm}

\subsection{Constraint Evolution}
\label{subsec:CEN}

\begin{figure*}[htbp]
\begin{minipage}[t]{.46\linewidth}
\includegraphics[angle=-90,width=6.5cm]{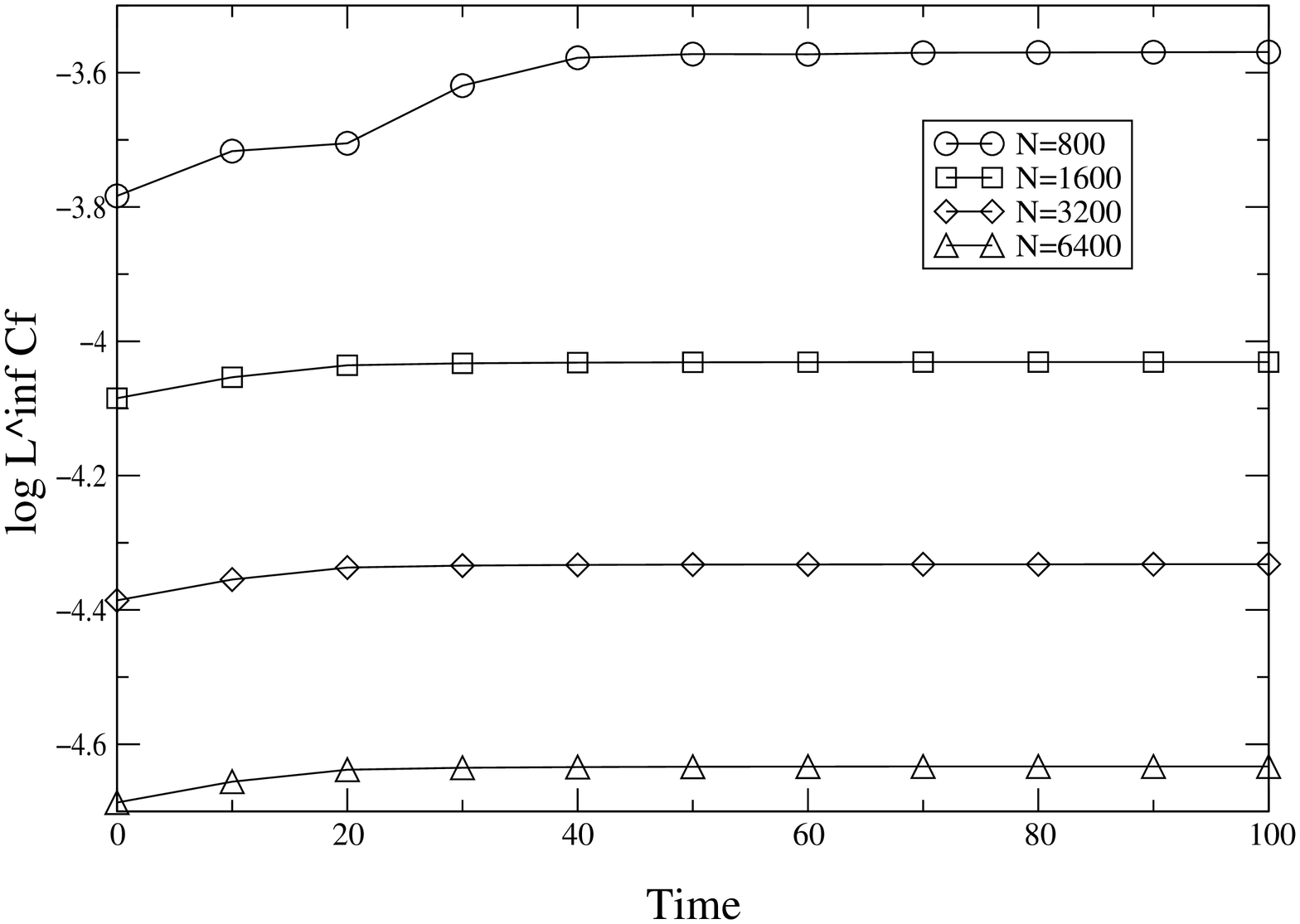}
\caption{Log of the $L^\infty$ norm of $C_f$ of type I data}
\label{fig:csI}
\end{minipage}
\hspace{2mm}
\begin{minipage}[t]{.46\linewidth}
\includegraphics[angle=-90,width=6.5cm]{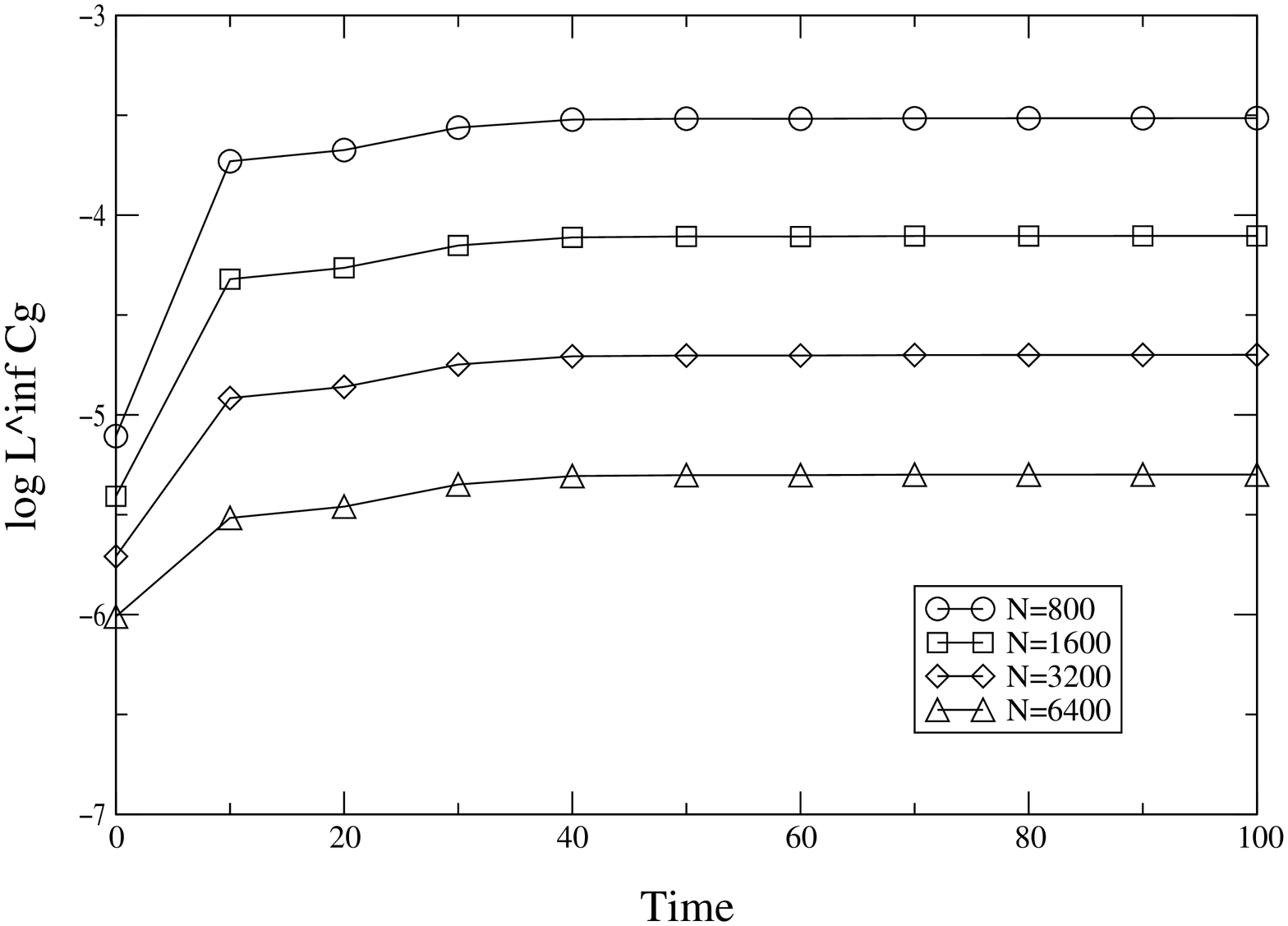}
\caption{Log of the $L^\infty$ norm of $C_g$ of type I data}
\label{fig:clI}
\end{minipage}

\vspace*{3mm}

\begin{minipage}[t]{.46\linewidth}
\includegraphics[angle=-90,width=6.5cm]{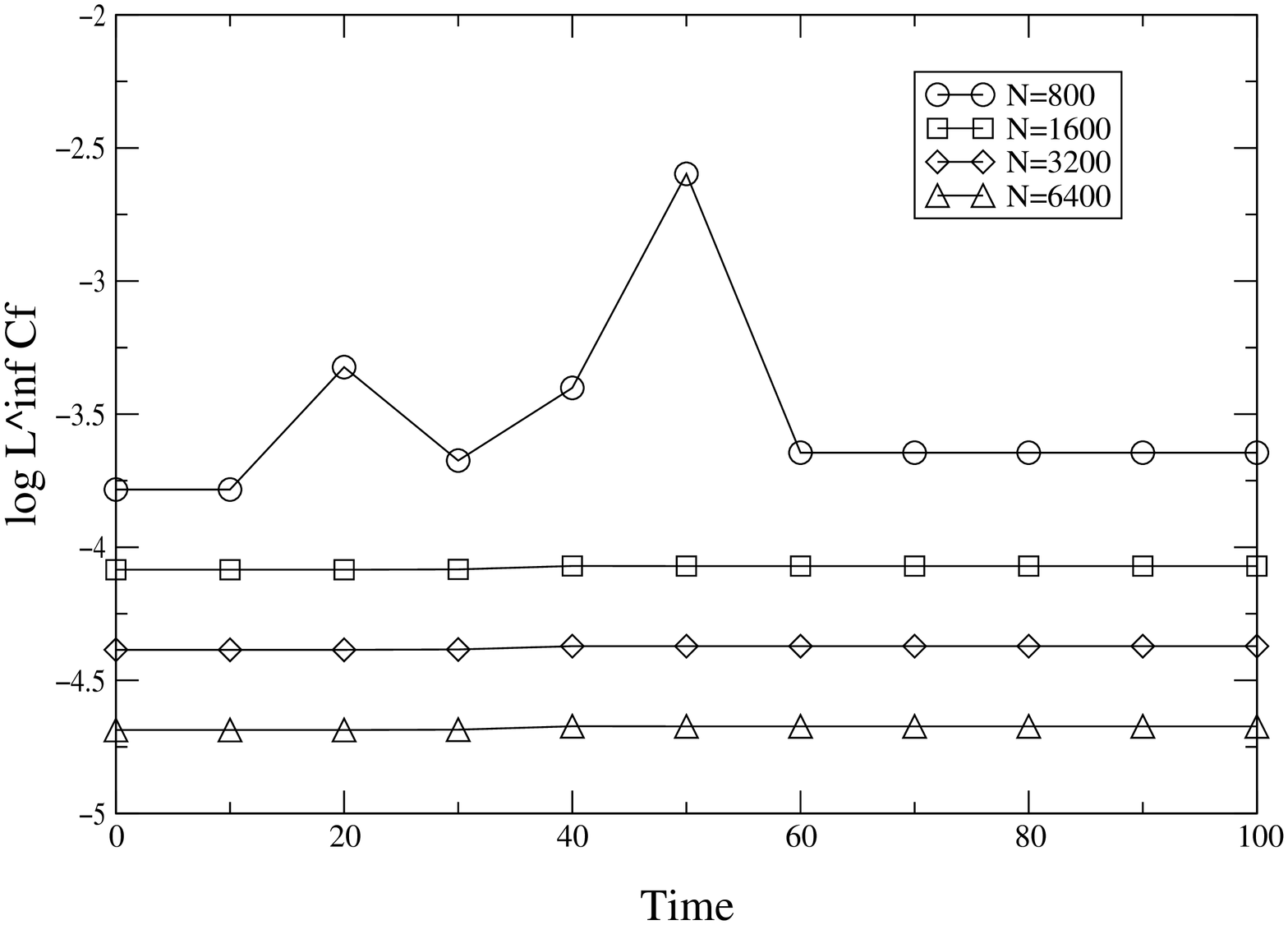}
\caption{Log of the $L^\infty$ norm of $C_f$ of type II data}
\label{fig:csII}
\end{minipage}
\hspace{2mm}
\begin{minipage}[t]{.46\linewidth}
\includegraphics[angle=-90,width=6.5cm]{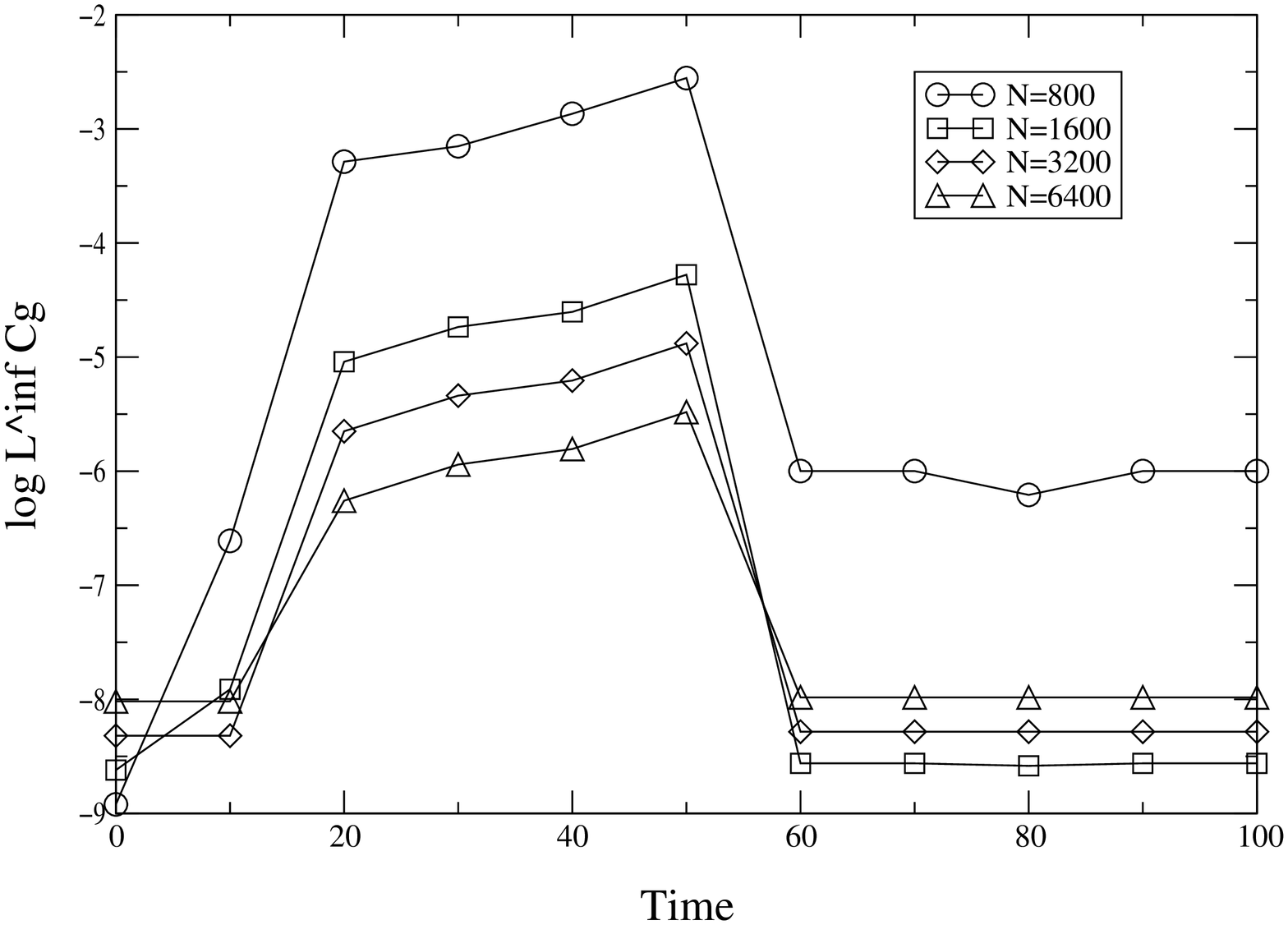}
\caption{Log of the $L^\infty$ norm of $C_g$ of type II data}
\label{fig:clII}
\end{minipage}

\vspace*{3mm}

\begin{minipage}[t]{.46\linewidth}
\includegraphics[angle=-90,width=6.5cm]{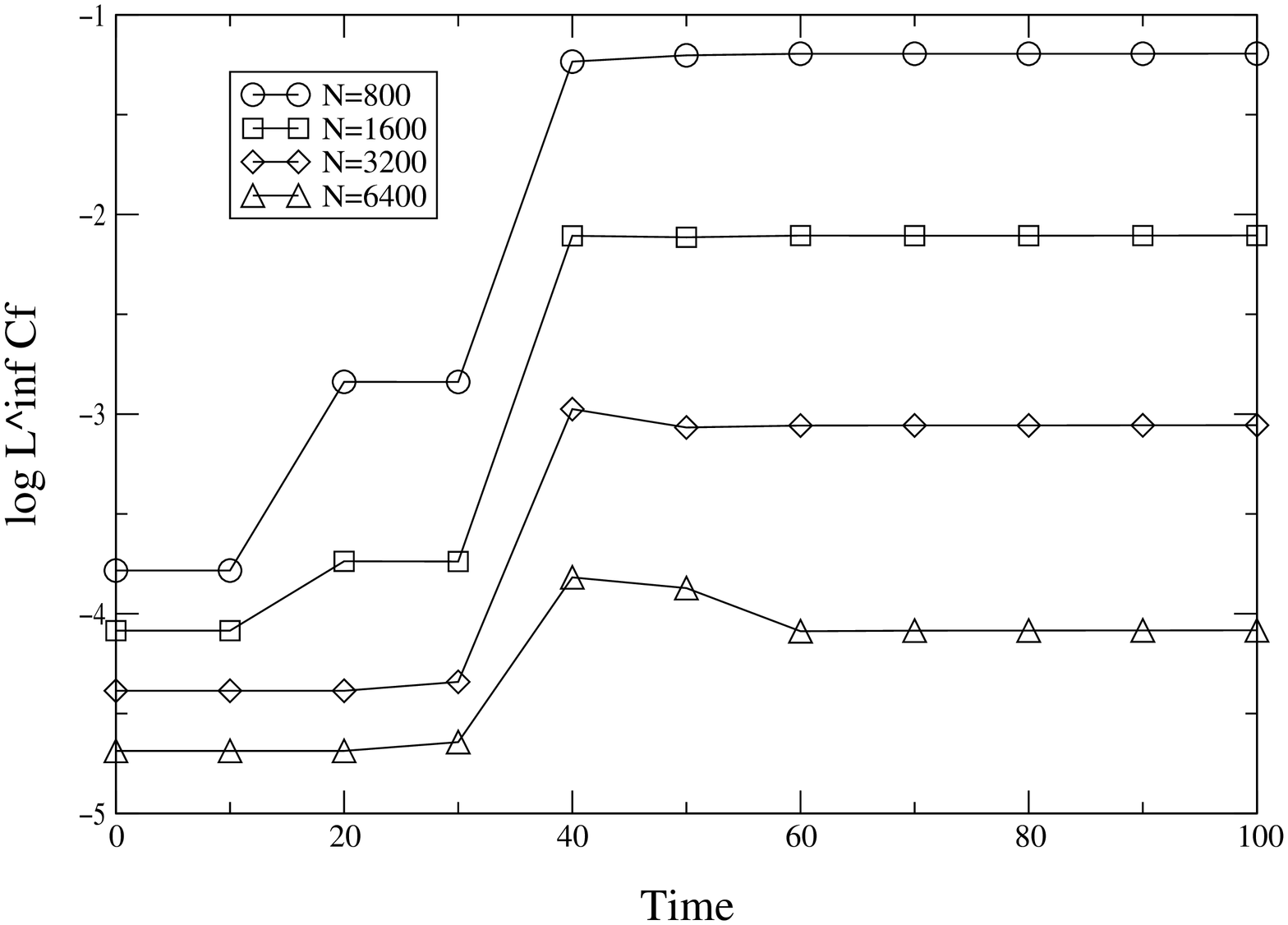}
\caption{Log of the $L^\infty$ norm of $C_f$ of type III data}
\label{fig:csIII}
\end{minipage}
\hspace{2mm}
\begin{minipage}[t]{.46\linewidth}
\includegraphics[angle=-90,width=6.5cm]{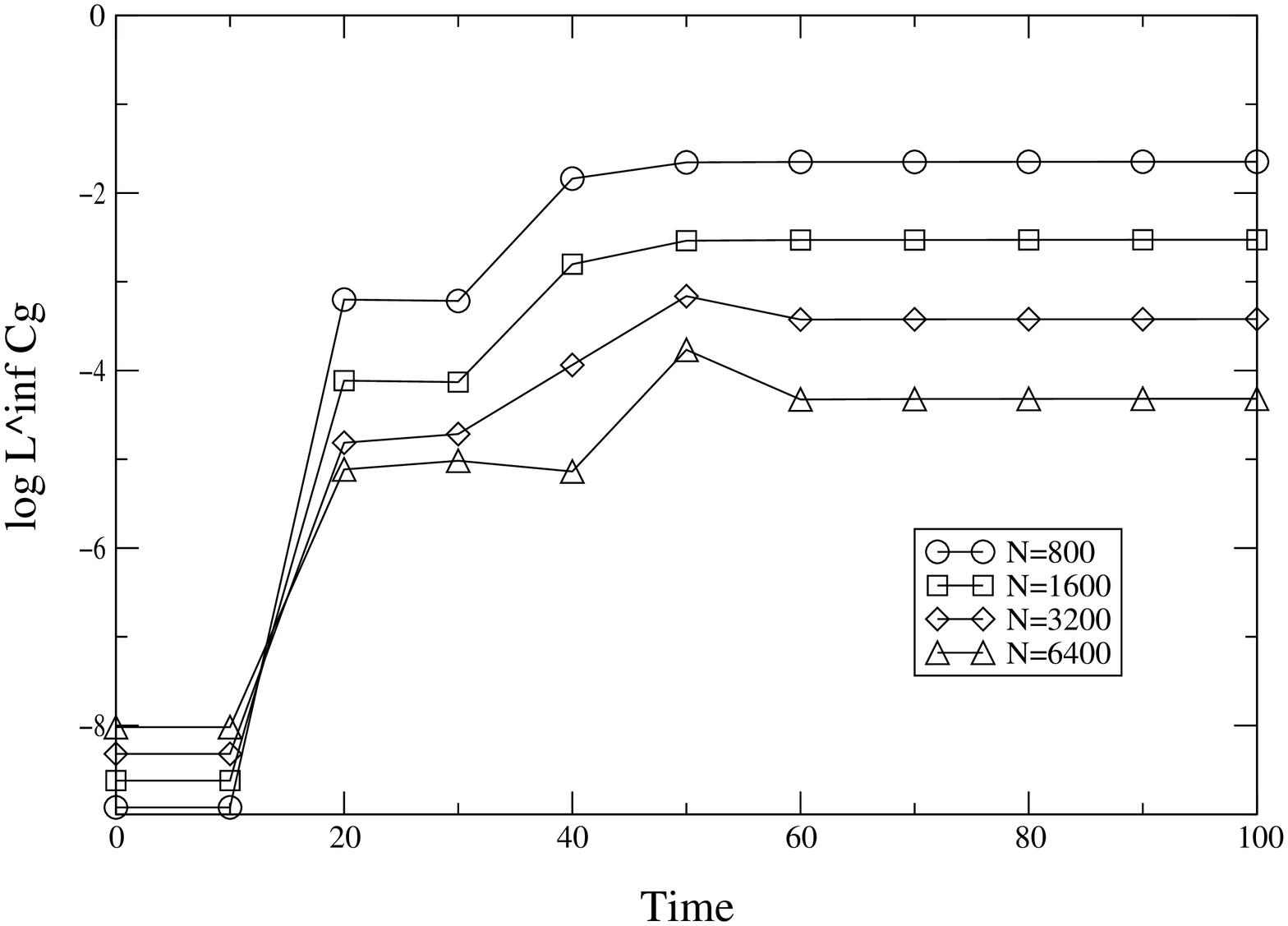}
\caption{Log of the $L^\infty$ norm of $C_g$ of type III data}
\label{fig:clIII}
\end{minipage}
\end{figure*}

The constraint equations are not solved for in any step of our simulations,
and although they are satisfied along evolution for the exact equations
here we have to check this happens at the discrete level. The figures 
(\ref{fig:csI}--\ref{fig:clIII}) show the $L^{\infty}$ and $L^2$ norm of the 
constraint expressions along evolution of different type of data set used. 
For type I data we had: $M_{Initial}=1$, $M_{Total}=1.15$, $M_{Final}=1.12$.
For the type II data we introduced at the outer boundary $10$ cycles of gauge 
mode of frequency $2\pi$, starting at $t=10$. The black hole mass was of course
constant and had the value $1$.
For the type III data we introduced at the outer boundary $10$ cycles of the
$\chi_+$ mode with frequency $2\pi$, starting at $t=10$. 
We had: $M_{Initial}=1$, $M_{Total}=2.6$, $M_{Final}=2.4$.


\vspace{2mm}


It is clear that the constraint quantities 
remain bounded along evolution and that they go to zero as the grid size
diminishes. Most of the contribution to the $L^{\infty}$ norm comes from
the boundary, where the derivative operator used is only second or third
order. These peaks near de boundary can be reduced to the same level than
the rest if one uses a discrete derivative operators which are fourth order 
at all grid points.

If one looks at the ratios $C_f/\frac{df}{dr}$, $C_g/\frac{dg}{dr}$ which
in some sense measure better the failure of the constraint to be satisfied
one sees that those ratios do not change so much along evolution.
For $C_f$ the change is of only about 7\%, and are of the order 
of $10^{-4}$.
For $C_g$, since $g$ is chosen initially to be zero, the change
at the beginning is rather big, but after a transient it gets to a plateau
of the order of $10^{-4}$.



\section{Results}
\label{sec:R}


\subsection{Decay of the scalar field, ringing and power law decay}

\begin{figure}[ht]
\begin{minipage}[t]{.46\linewidth}
\centering \includegraphics[angle=-90,width=8cm]{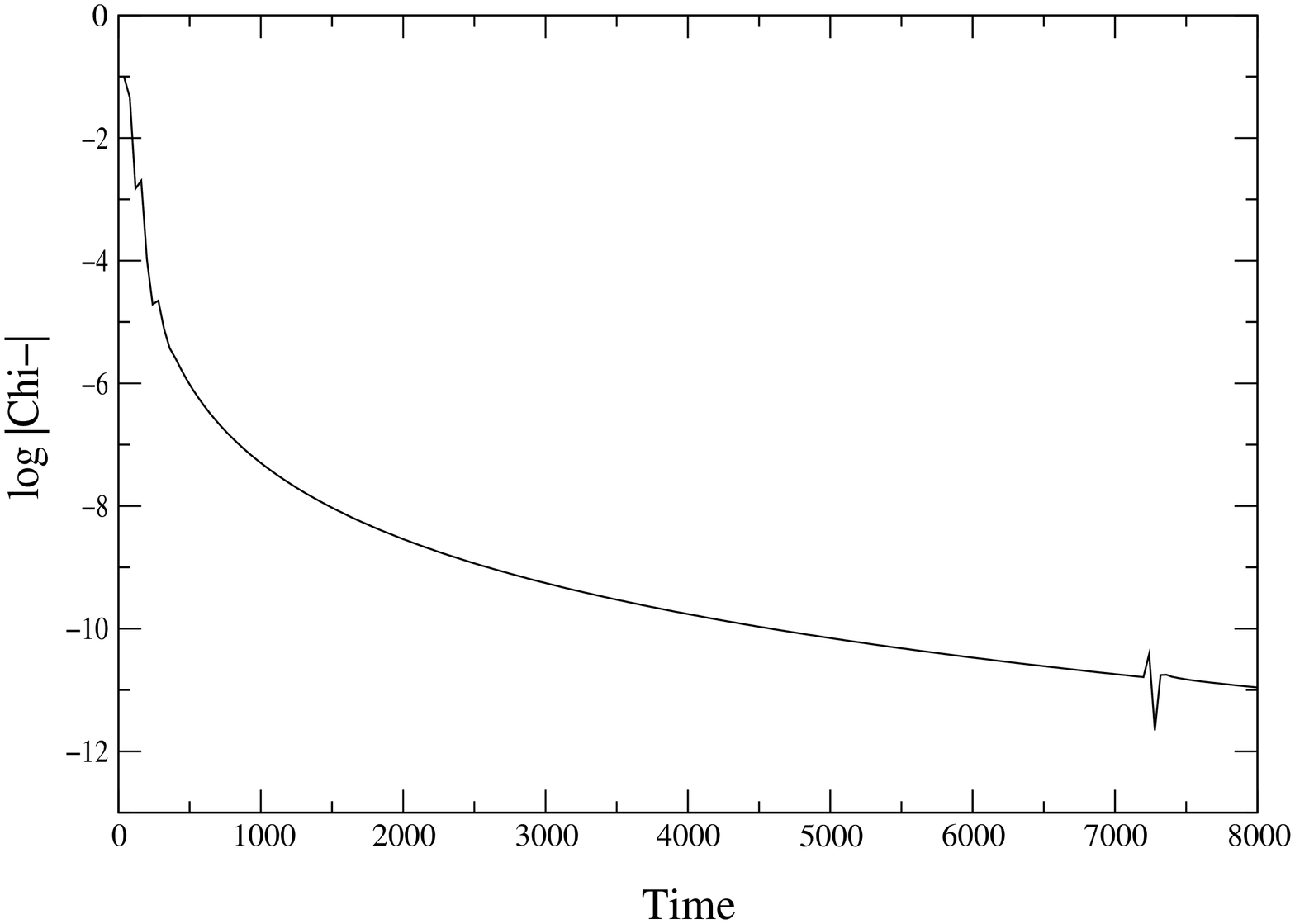}
\caption{Decay of $\chi_-$ at $r=25.5$}
\label{fig:decay1}
\end{minipage}
\hspace{2mm}
\begin{minipage}[t]{.46\linewidth}
\centering \includegraphics[angle=-90,width=8cm]{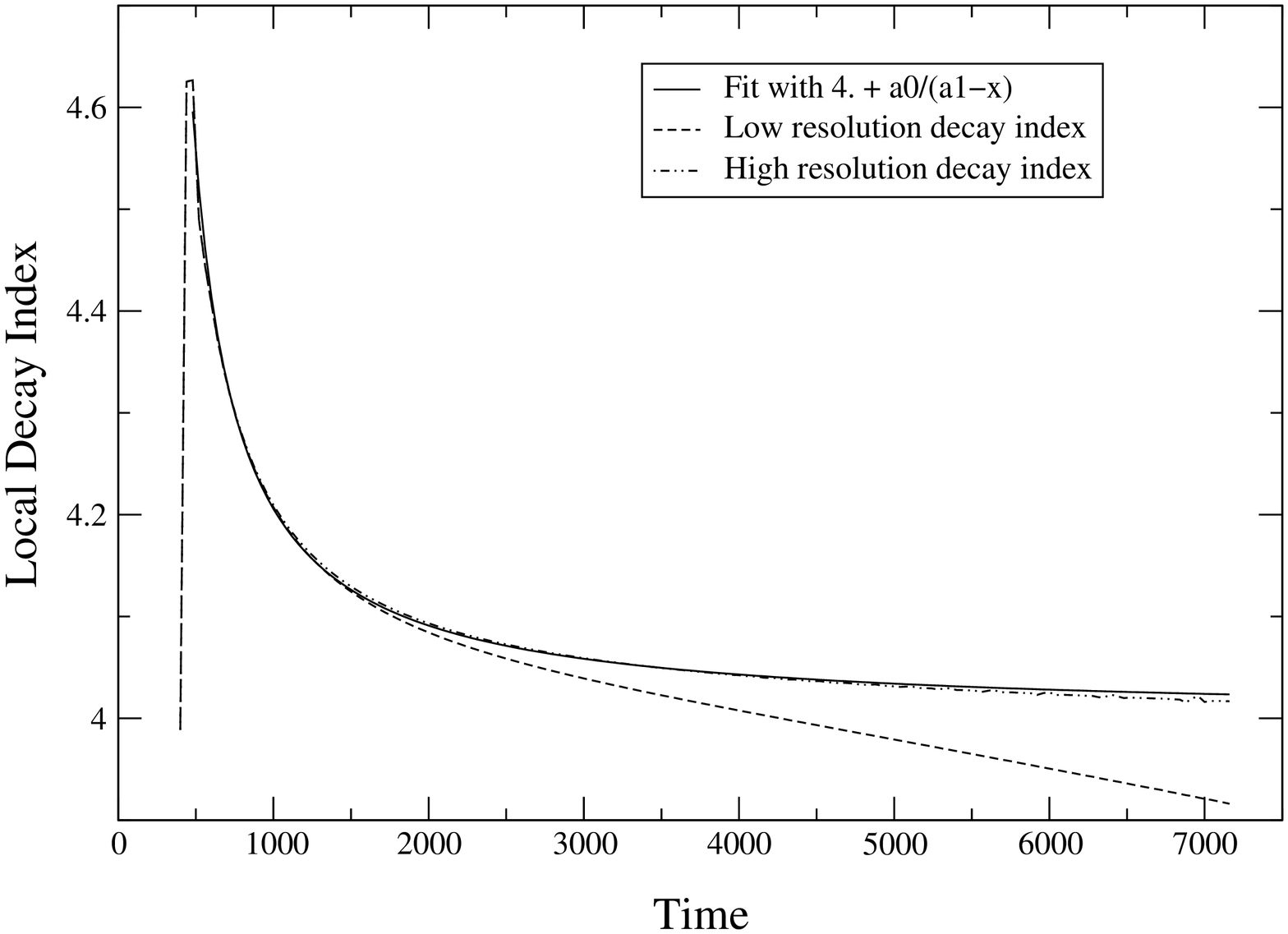}
\caption{Power law decay of $\chi_-$ at $r=25.5$}
\label{fig:decay2}
\end{minipage}
\end{figure}

We reproduce standard results on ringing and tail (power law decay)
for the first type of simulation.
In figure (\ref{fig:decay1}) it is plotted the 
log of the absolute value of $\chi_-$ at $r=25.5$ with respect to time. 
It is a run where the outer boundary is at $R_{outer}=6000$, the space grid
spacing was $h/M = 0.0234$, where we have used the initial mass to make the
ratio.~footnote{In most of the runs $k/h \approx 2$, this does not violate the
Courant-Friedrichs-Lewi condition because we use Runge-Kutta for time evolution.}  
The initial scalar field is given by a Gaussian $\chi_+$ pulse centered at 
$r=10$, giving: Initial black hole mass $M_{Initial}=1$, total mass 
$M_{Total} = 2.56$, final black hole mass $M_{Final}=2.45$.  
At the beginning the characteristic ringing is seen and then a
power law decay of the type $t^{\alpha}$ 
(this can be seen best plotting the local decay power 
$\displaystyle{t \frac{d\ln(\chi_-)}{dt}}$
as defined in \cite{Burko_1,Burko_2}, 
figure (\ref{fig:decay2})). 
The power law decay computed
$t^{-4.01}$ (at $t=7200$), agrees quite well from the one 
expected for linear perturbation theory, namely $t^{-4}$.
See \cite{GPP_1,GPP_2}. 
In the plot we also show a lower resolution ($h/M = 0.0468$) computation
where it is clear that an instability starts to appear and become important 
at about half the time. 
This instability diminishes when the grid resolution is incremented,
but still seems to be present at longer times. 
At the resolution used a feature departing form the power law 
decay is seeing  after $t=7200$, this corresponds to an higher
frequency mode propagating backwards at a speed $1.66$ the speed of light. 
These high frequency modes are common on all centered fourth order methods.
Their amplitude decreases with increased precision.
The time of arrival of that feature implies it originates near the inner 
boundary probably due to imperfect ingoing boundary conditions.
After that feature, we again see (not shown in the figure) a even
bigger feature, this one propagates at the correct speed of the problem, but
its presence at that time implies that it is generated by the one traveling at
higher speed when it enters the region near the outer boundary.

\vspace{2mm}

The possibility of giving boundary conditions which would automatically 
satisfy the constraint equations allow as to perform several interesting
runs. In particular we made a couple of runs of type III with 
$$
\chi_+|_{R_{out}} = A_+ (1 - \cos(\omega_+ t)) \qquad t \in [10,20]
$$ 
with $A_+ = 1/2\pi$, and  in one case $\omega = 2\pi$, in the other $\omega = 2\pi/10$. For these two boundary data we have,

$$
\int^{20}_{10} \chi_+^2 dt = 15A_+^2
$$

\begin{figure}[ht]
\centering \includegraphics[width=8cm]{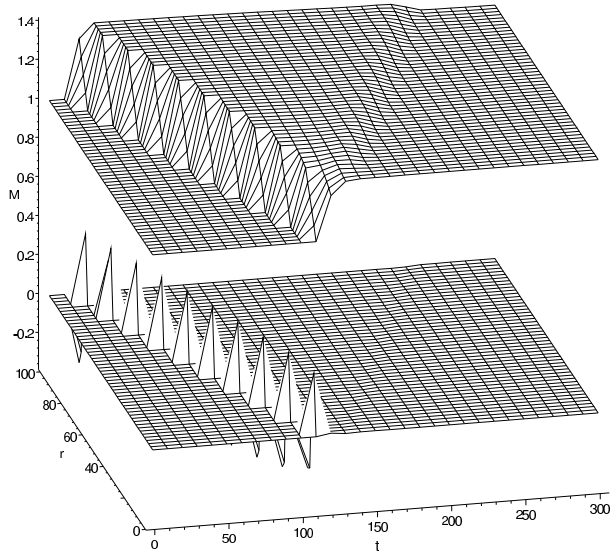}
\caption{Mass of type III run and mass difference between runs augmented 5 times.}
\label{fig:mass_71}
\end{figure}

\begin{figure}[ht]
\centering \includegraphics[angle=-90,width=8cm]{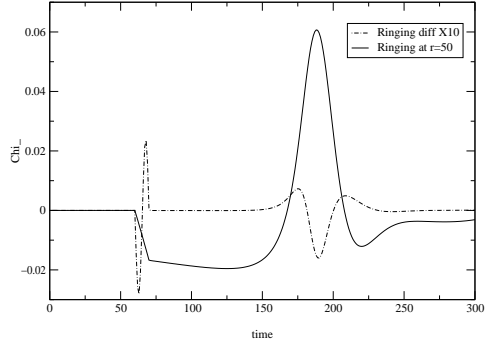}
\caption{Values of $\chi_-$ at $r=50$ for a type III run, bellow the difference between runs
augmented 10 times.}
\label{fig:ring_71}
\end{figure}

Since the energy flux is given by 

$$
\frac{dm}{dt} = \frac{1}{g}(\chi_+^2 - \frac{(f-2)^2}{f^2}\chi_-^2),
$$
and $\frac{dg}{dt} \approx \frac{2g}{fr}\chi_+^2$,  
$\frac{df}{dt} \approx \frac{4}{fr}\chi_+^2$, 
$\frac{d\chi_-}{dt} \approx \frac{-1}{r}\chi_+$.
We see that to a 1\% accuracy the flux of both solutions should coincide
and so the black hole masses should be similar. Even more, the ringing
of both solutions should be very similar, for it mostly depends on the
geometry.
In figure (\ref{fig:mass_71}) we see the mass of one of the simulations and
the mass difference between both simulations multiplied by 5. We see that
most of the difference is bounded to the region where the wave of scalar
field is moving, leaving behind the same geometry.
In figure (\ref{fig:ring_71}) we see one of the ringings (the value of $\chi_-$
at $r=50$) and the difference between the ringing of both solutions augmented
by a factor 10. Again we see that most of the difference is at the moment
where the two different wave fronts pass the point and then on the precise
location of the maxima of the ringing.

\subsection{Relation between the initial mass and the black hole mass}
\label{subsec:M}

\begin{figure}[ht]
\begin{minipage}[t]{.46\linewidth}
\centering \includegraphics[angle=-90,width=8cm]{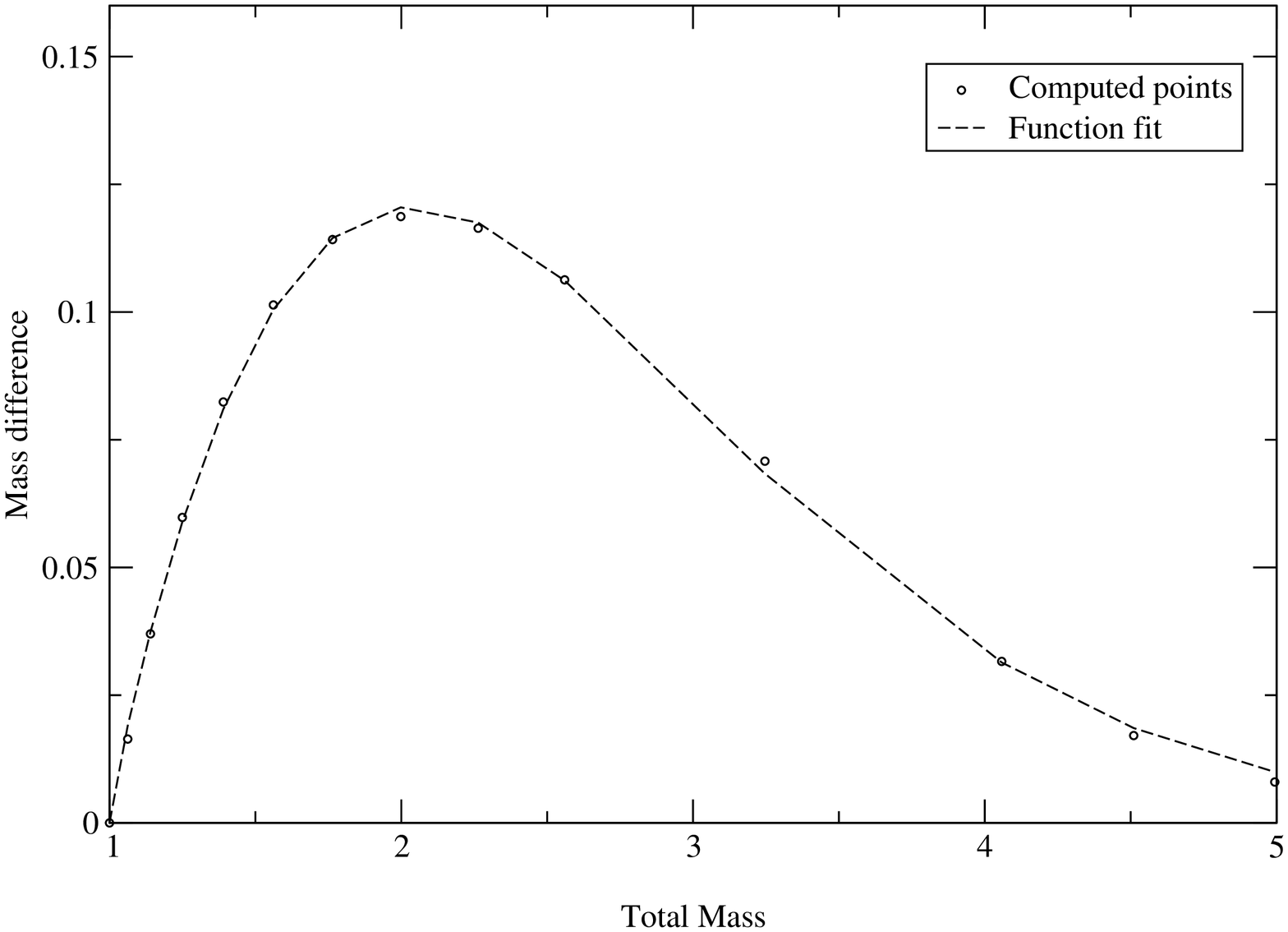}
\caption{$\Delta M$ vs $M$ for type I data, $\Delta M = 0.21*(M-1)^{0.86}\;e^{-0.55(M-1)^{1.47}}$}
\label{fig:mass_1}
\end{minipage}
\hspace{2mm}
\begin{minipage}[t]{.46\linewidth}
\centering \includegraphics[angle=-90,width=8cm]{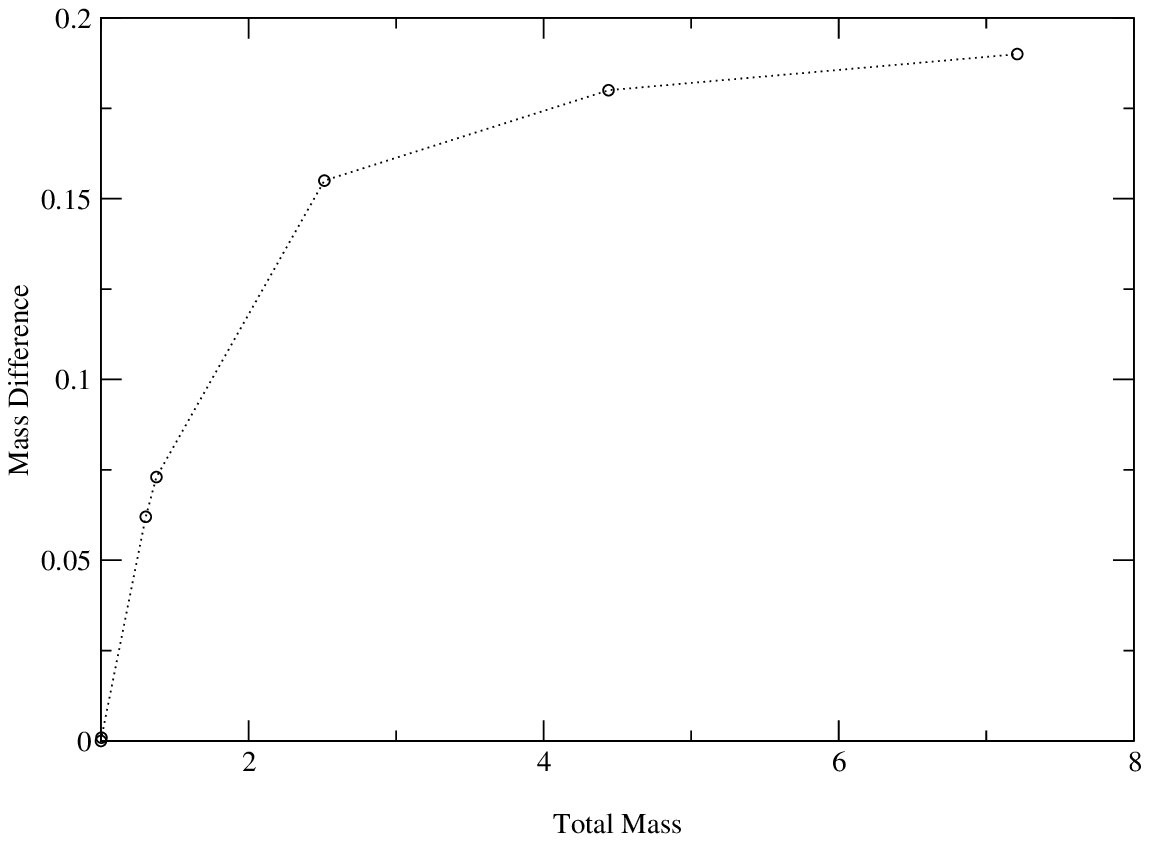}
\caption{$\Delta M$ vs $M$ for type III data}
\label{fig:mass_2}
\end{minipage}
\end{figure}

We have tested the relation between the mass gap given by the total initial 
mass minus the final black hole mass as a function of the total initial mass
for type I and type III initial-boundary data sets. 
In figures (\ref{fig:mass_1}, \ref{fig:mass_2})
we display the results. In both cases we start with a black hole of mass unit,
and put our inner boundary inside its horizon. 

\section{Conclusions}
\label{sec:C}


It is difficult to see whether some of the ideas presented in this work
can be extended to the full three dimensional case, where many more 
variables are present and where there is no preferred center in which
to anchor a gauge as the one used here. But the model is so simple that 
perhaps the observations made in solving it can shed some light into the 
more difficult problem. Among the observations we have the following:
\textit{a)} If the gauge prescription does not fix completely the gauge
it is expected there would be gauge modes propagating. In this case there
was just one gauge freedom left (the value of $h$ at the initial surface
and at the outer boundary), nevertheless there were three equations which
acquired a non trivial propagation.
\textit{b)} Gauge modes, by being non-physical, can have singular behavior
although the geometry given by the solution can be regular. It is clear 
in the model studied that the different components of the metric grow
exponentially while the solution is still Schwarzschild.
Thus, it seems to be necessary to isolate and keep under control the potentially
unstable gauge modes. In our case this was done by choosing variables for 
which it was clear where the instability was, and so, choosing conveniently
the initial-boundary data we could keep it small for many cases, although
not always.
\textit{c)} The choice of gauge modes can also complicate the boundary
value problem. Ideally some of the fields should acquire values at the
boundary so that the constraint propagate correctly. In the model it can
be seen that if part of the data for the pair $(f,g)$ must be given at
the boundary (according to the values chosen for the coefficients 
$(K_{ff},K_{fg},K_{gg},K_{gf})$) that part must satisfy some evolution
equation along the boundary, so actually there is no freedom left.
But other gauge quantities, like $h$, can take any value. Thus, if the
variables are not chosen appropriately it could become very cumbersome
to find the correct boundary conditions which would, at the same time
keep the constraint equations being satisfied and the gauge instabilities
under control. Note that the procedure we used, getting some equations which
are intrinsic to the boundary for some of the variables, is similar to the 
one used in \cite{F_N}, the only case where well posedness of the 
initial-boundary value problem in full general relativity has been asserted.
%



\section{Acknowledgments}
\label{sec:A}

It is a pleasure to acknowledge several and important discussions 
with S. Frittelli,  
H.O. Kreiss, P. Laguna, and M. Tiglio.


\end{document}

%% file: char.pstex_t
\begin{picture}(0,0)%
\epsfig{file=char.pstex}%
\end{picture}%
\setlength{\unitlength}{2486sp}%
\begingroup\makeatletter\ifx\SetFigFont\undefined%
\gdef\SetFigFont#1#2#3#4#5{%
  \reset@font\fontsize{#1}{#2pt}%
  \fontfamily{#3}\fontseries{#4}\fontshape{#5}%
  \selectfont}%
\fi\endgroup%
\begin{picture}(6357,4872)(856,-4921)
\put(1846,-916){\makebox(0,0)[lb]{\smash{\SetFigFont{8}{9.6}{\rmdefault}{\mddefault}{\updefault}$(f,g)$}}}
\put(6211,-4111){\makebox(0,0)[lb]{\smash{\SetFigFont{8}{9.6}{\rmdefault}{\mddefault}{\updefault}$(f,g)$}}}
\put(3151,-2986){\makebox(0,0)[lb]{\smash{\SetFigFont{8}{9.6}{\rmdefault}{\mddefault}{\updefault}$(h,\chi_+)$}}}
\put(4951,-2536){\makebox(0,0)[lb]{\smash{\SetFigFont{8}{9.6}{\rmdefault}{\mddefault}{\updefault}$\chi_-$}}}
\put(1171,-2086){\makebox(0,0)[lb]{\smash{\SetFigFont{8}{9.6}{\rmdefault}{\mddefault}{\updefault}$\chi_-$}}}
\put(3106,-1636){\makebox(0,0)[lb]{\smash{\SetFigFont{8}{9.6}{\rmdefault}{\mddefault}{\updefault}horizon ($f=2$)}}}
\put(856,-4876){\makebox(0,0)[lb]{\smash{\SetFigFont{8}{9.6}{\rmdefault}{\mddefault}{\updefault}$R_{in}$}}}
\put(6841,-4921){\makebox(0,0)[lb]{\smash{\SetFigFont{8}{9.6}{\rmdefault}{\mddefault}{\updefault}$R_{out}$}}}
\put(4951,-466){\makebox(0,0)[lb]{\smash{\SetFigFont{8}{9.6}{\rmdefault}{\mddefault}{\updefault}$(h,\chi_+)$}}}
\end{picture}